\documentclass[preprint,review,12pt,sort&compress]{elsarticle}

\usepackage{graphicx}
\usepackage{dcolumn}
\usepackage{bm}
\usepackage{multirow}
\usepackage{hyperref}
\usepackage{booktabs}

\hypersetup{
colorlinks=true,
linkcolor=blue,
citecolor=blue,
urlcolor=blue
}

\usepackage{chemformula}
\usepackage[output-decimal-marker={.}]{siunitx}

\DeclareSIUnit\angstrom{\text{\AA}}

\journal{Applied Surface Science}

\begin{document}

\begin{frontmatter}



\title{Phosphorus-based lubricant additives on iron with Machine Learning Interatomic Potentials} 


\author[inst1]{Paolo Restuccia\corref{contrib}}

\affiliation[inst1]{organization={Dipartimento di Fisica e Astronomia, Università di Bologna},
            addressline={Viale Berti Pichat 6/2}, 
            city={Bologna},
            postcode={40127}, 
            country={Italy}}

\author[inst1]{Enrico Pedretti\corref{contrib}}

\author[inst1]{Francesca Benini}

\author[inst2]{Sophie Loehlé}

\affiliation[inst2]{organization={TotalEnergies, OneTech Fuels\&Lubricants, Research Center Solaize},
            addressline={Chemin du Canal BP 22}, 
            city={Solaize},
            postcode={69360}, 
            country={France}}

\author[inst1]{M. Clelia Righi\corref{mycorrespondingauthor}}
\ead{clelia.righi@unibo.it}
\cortext[mycorrespondingauthor]{Corresponding author}
\cortext[contrib]{Authors contributed equally}

\begin{abstract}
Phosphorus-based lubricant additives are used for protecting metallic contacts under boundary lubrication by forming surface films that reduce wear and friction. Despite their importance, the molecular mechanisms driving their friction-reducing effects remain unclear, especially for phosphate esters, whose molecular structure critically impact tribological behavior. In this study, we use machine learning–based molecular dynamics simulations to investigate the tribological performance of three representative phosphorus-based additives, Dibutyl Hydrogen Phosphite (DBHP), Octyl Acid Phosphate (OAP), and Methyl Polyethylene Glycol Phosphate (mPEG-P), on iron surfaces. The mPEG-P family is further analyzed by varying esterification degree and chain length. DBHP exhibits the lowest friction and largest interfacial separation, resulting from steric hindrance and tribochemical reactivity, as indicated by P--O bond cleavage and enhanced O--Fe interactions. In contrast, OAP and mPEG-P monoesters produce higher friction due to limited steric protection and reduced resistance to shear, leading to partial loss of surface coverage under extreme conditions. Within the mPEG-P family, multi-ester and longer-chain molecules significantly lower friction by maintaining larger separations, demonstrating that steric effects can outweigh surface reactivity under severe confinement. Overall, these results provide atomistic insights into how molecular architecture controls additive performance and support the design of phosphorus-based lubricants combining reactive anchoring with optimized steric structures for durable, low-friction interfaces.
\end{abstract}


\end{frontmatter}



\section{Introduction}
\label{sec:intro}

Lubrication is a fundamental process that enables the reliable operation of mechanical contacts, where friction and wear must be minimized to ensure long-term performance and durability. Among the different lubrication materials, oils containing specific chemical additives are the most widely employed in engineering and industrial applications. These additives are essential components of lubricant formulations, providing functionalities, such as anti-wear, anti-oxidant, and anti-corrosive protection, that base oils alone cannot offer. When the liquid film becomes too thin to sustain the applied load, many sliding components operate under boundary or mixed lubrication regimes. In such conditions, direct asperity contacts can occur, and lubricant additives become the primary agents controlling friction and wear. These molecules adsorb onto the metallic surfaces and undergo mechanochemical reactions that generate protective tribofilms, preventing metal-to-metal contact and reducing interfacial shear~\cite{Spikes_2008,Tang_2014,Furlong_2007}. Designing advanced additives capable of efficiently performing these functions remains a central challenge in tribology.

Among the wide range of lubricant additives, phosphorus-containing compounds, such as phosphate esters, phosphites, and combined phosphorus/sulphur species, have long been recognized as indispensable ingredients in boundary lubrication. Their remarkable anti-wear and extreme-pressure performance arises from their ability to form protective surface films that simultaneously minimize wear and adjust frictional properties~\cite{Johnson_16,Emmanuel_2024}. The underlying tribochemical reactions are complex and typically yield a mixture of iron phosphates, polyphosphates, and phosphides, which together protect the interface and lower shear stresses under demanding conditions~\cite{DeBarros-Bouchet_2015}. As a prerequisite to these processes, tribofilm formation is typically initiated by the adsorption of lubricant additives at the metal surface. In the present work, we focus on elucidating these early-stage adsorption mechanisms for phosphorus-based lubricant additives.

Within this chemical family, phosphate esters represent a particularly versatile class whose molecular architecture critically determines their tribological performance. Factors such as alkyl or aryl substitution, chain length, branching, and degree of esterification all influence how these compounds adsorb, react, and decompose under sliding conditions~\cite{Yang_2024,Wu_2016}. To unravel how molecular structure influences tribological behaviour, three representative phosphorus-based additives are considered: Octyl Acid Phosphate (OAP), Dibutyl Hydrogen Phosphite (DBHP), and Methyl Polyethylene Glycol Phosphate (MPEG-P) whose structures are shown in Fig.~\ref{fig:mols}. Each embodies a distinct structural motif and provides complementary insights into the molecular origins of lubricant performance.

OAP is a monoester phosphate with a relatively long alkyl chain, striking a balance between surface activity and hydrophobicity. Its asymmetric structure, comprising a single octyl group bound to the phosphate group and two acidic hydrogens, enhances both hydrogen bonding and surface reactivity, shaping its adsorption behaviour and the nature of the resulting tribofilm~\cite{Johnson_16}. DBHP, in contrast, is a phosphite additive characterized by a trivalent phosphorus center with a P=O double bond. This structural feature leads to distinct tribochemical pathways, as phosphites can transform into phosphorus-rich species such as iron phosphides, which are particularly efficient in lowering interfacial shear~\cite{DeBarros-Bouchet_2015}. MPEG-P, instead, integrates a phosphate ester headgroup with a polyethylene glycol (PEG) tail terminated by a methyl group, producing an amphiphilic molecule that combines hydrophilic PEG chains with hydrophobic termination~\cite{Penczek_2009}. Its synthesis typically involves phosphorylation of methyl-terminated PEG followed by hydrolysis or neutralization, yielding mixtures of mono-, di-, and triesters with varying amphiphilicity. The degree of esterification governs the surface affinity: monoesters (RO–PO$_3$H$_2$) exhibit strong acidity and adsorption on metal oxides, diesters ((RO)$_2$–PO$_2$H) maintain surface activity with greater hydrophobicity, and triesters ((RO)$_3$–PO) are more soluble in non-polar lubricants but less surface-reactive.

Despite extensive experimental research, a complete understanding of how molecular architecture controls the balance between reactivity, steric effects, and film-forming ability remains elusive. Computational methods, particularly molecular dynamics (MD), have become invaluable in exploring such molecular-scale mechanisms. Recent MD studies have shown that even subtle structural modifications can alter decomposition kinetics and tribofilm composition. For example, branched alkyl substituents accelerate mechanochemical breakdown relative to linear analogues, with tri(s-butyl) phosphate decomposing faster than tri(n-butyl) phosphate owing to larger pre-exponential factors in stress-temperature relationships~\cite{Latorre_2021}. Moreover, the substrate composition, like iron versus iron oxide, strongly affects reaction pathways and final tribofilm chemistry~\cite{Yang_2024}.

However, conventional classical MD is limited by its inability to capture bond-breaking and electronic effects with quantum accuracy. The recent emergence of machine-learning interatomic potentials (MLPs) bridges this gap, combining near-first-principles accuracy with the efficiency of classical MD. These approaches now allow systematic investigation of how chain length, esterification degree, and steric hindrance govern interfacial reactivity and frictional response under realistic tribological conditions~\cite{ref:Pacini_2024,ref:Ta_2025}.

In this work, we leverage these advances to perform a comparative study of DBHP, OAP, and MPEG-P using machine-learning-based molecular dynamics simulations. By correlating molecular structure with frictional behaviour, interfacial separation, and surface chemistry, we aim to uncover the fundamental design principles governing the performance of phosphorus-based lubricant additives. The insights gained from this study are expected to inform the rational development of next-generation lubricant formulations with improved performance and durability.

\section{Methods}
\label{sec:methods}

\subsection{Training machine learning interatomic potentials}

\begin{figure}[htpb]
\includegraphics[width=\columnwidth]{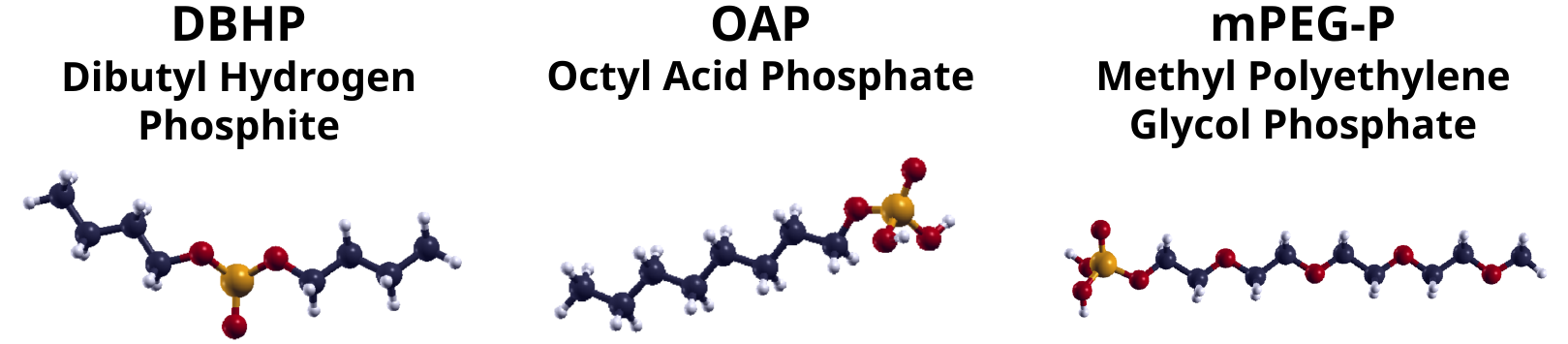}

\vspace{0.75cm}

\includegraphics[width=\columnwidth]{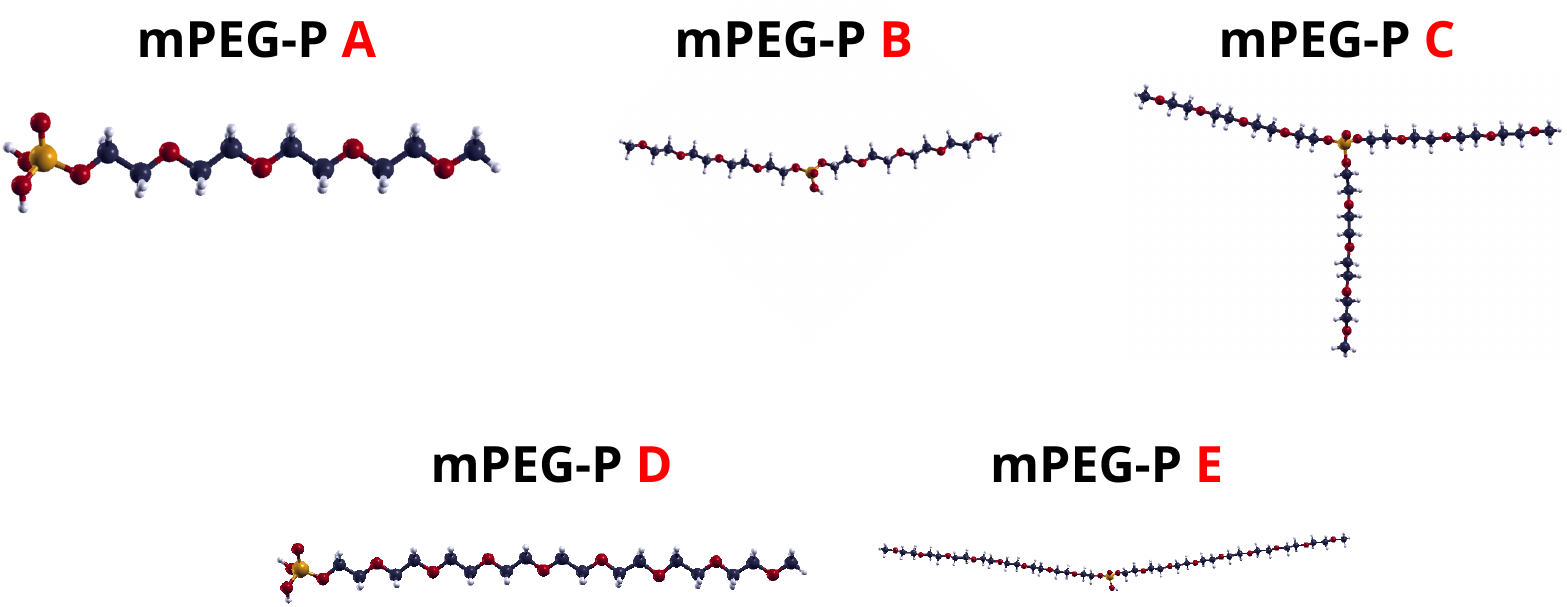}

\caption{\label{fig:mols} Ball-and-stick representation of the molecules studied in this work. From now on, white atoms represent H, grey C, yellow P and red O, respectively. In the lower panels, mPEG-P molecules with different number of esters and glycol units are shown. We modelled mPEG-P monoester with 4 glycol units (label A in red), mPEG-P diester with 4 glycol units (label B), mPEG-P triester with 4 glycol units (label C), mPEG-P monoester with 8 glycol units (label D), and mPEG-P diester with 8 glycol units (label E).\label{fig:esters}}
\end{figure}

To perform large scale molecular dynamics simulations of tribological systems while preserving high accuracy in the description of chemical bonding, it is necessary to develop machine learning interatomic potentials (MLIPs) trained on DFT data. This task is inherently challenging, as it requires sampling a broad range of atomic configurations to capture the extreme conditions typically encountered by lubricant mixtures and tribological interfaces in sliding nano-contacts. The accuracy of a MLIP critically depends on the variety of the underlying DFT dataset, making it paramount to identify and include the most relevant conditions across a wide range of the configurational space. To address this issue, we adopted an active learning strategy using the Smart Configuration Sampling (SCS) framework developed by our group~\cite{ref:Pacini_2025}. SCS enables efficient exploration of the configuration space by performing explorative molecular dynamics in tribological conditions using the MLIP. This enables the simulation of much longer time scales than those accessible by ab initio molecular dynamics, allowing to explore a wider variety of local structures and to capture rare events. Through an iterative process, it expands the \textit{ab initio} dataset by identifying the most relevant atomic configurations encountered during the dynamics and computing their associated energies and forces.

\begin{figure}[htpb] 
    \centering 
    \includegraphics[width=0.7\columnwidth]{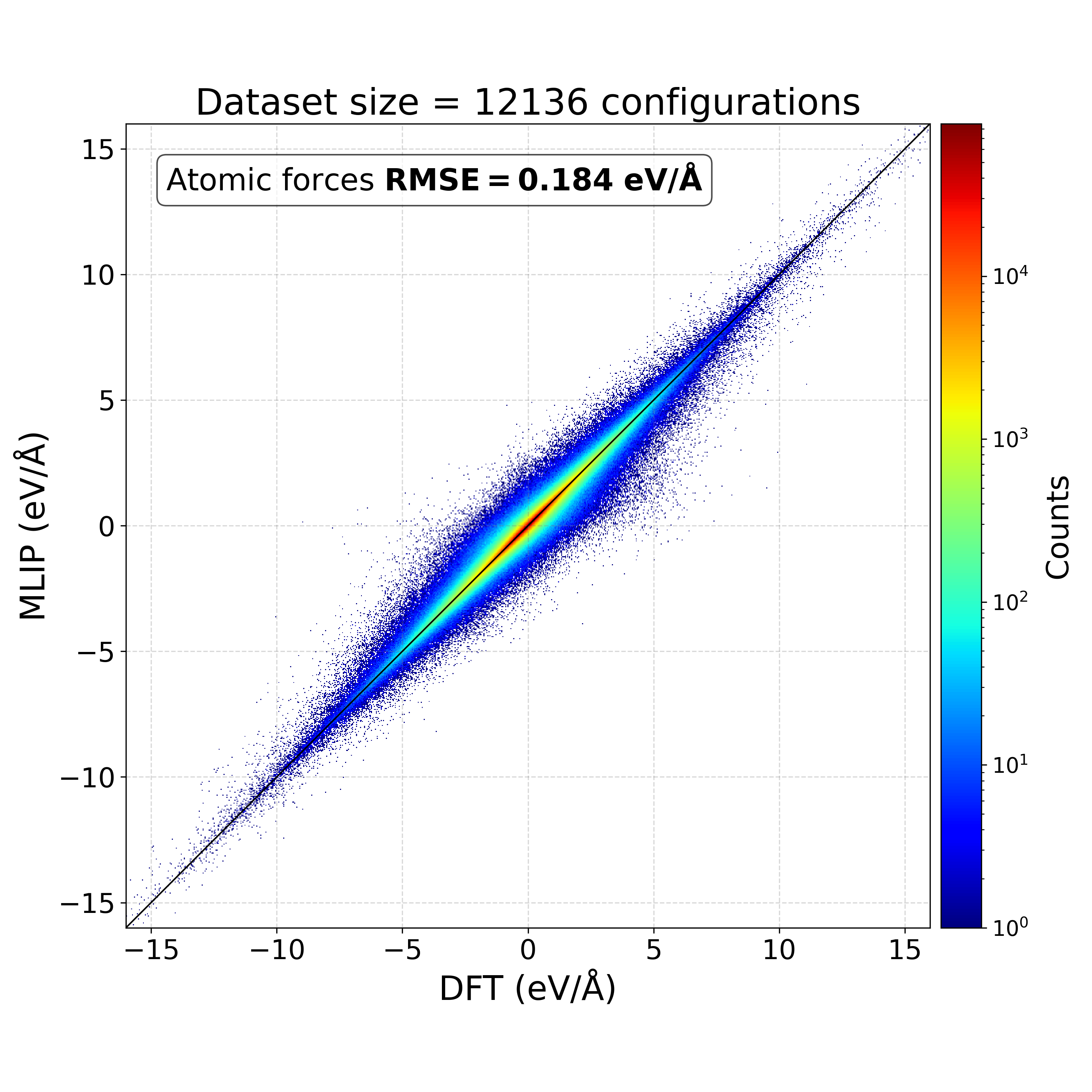}
    \caption{Parity plot for the DeePMD model evaluated on the complete dataset, comparing predicted atomic forces from the MLIP (y-axis) with reference DFT values. (x-axis). The colour scale indicates the density of data points, with warmer (cooler) colours representing higher (lower) counts of occurrences at specific force values. The overall RMSE is also reported in the plot.} 
    \label{fig:par_plot} 
\end{figure}

The active learning procedure led to the identification of approximately 12136 \textit{ab initio} configurations, which formed the final dataset used to train the MLIP. For the MLIP model we employed the DeePMD-kit~\cite{ref:DeePMD-v2} architecture, using the two-body embedding DeepPot-SE (\texttt{se\_e2\_a}) descriptor with [25, 50, 100] neurons, and a fitting net containing [120, 120, 120] neurons. Fig.~\ref{fig:par_plot} shows the model's accuracy in predicting atomic forces on the whole dataset (whose systems are detailed in Section~\ref{sec:mol_system}), achieving a root mean square error (RMSE) of 0.18 eV/\r{A}. This level of accuracy is consistent with state-of-the-art MLIPs: recent benchmarks demonstrate that RMSEs of this magnitude enable accurate simulations of complex systems while maintaining computational efficiency~\cite{ref:Liang_2023}. Moreover, error analyses confirm that force RMSEs in the range of 0.15–0.4 eV/\r{A} are widely considered acceptable for reliable predictive modelling~\cite{ref:Liu_2023}.

\subsection{Set up \textit{ab initio} calculations}

All the \textit{ab initio} calculations were performed using spin-polarised DFT as implemented in the GPU-enabled version of the Quantum Espresso computational suite~\cite{ref:QE_GPU}. We adopted the same computational parameters for all the considered systems, following standard practice in MLIP development. To ensure these shared settings yielded reliable results across the different active learning simulations, we used sufficiently large simulation cells to minimise discrepancies between systems. In particular, the general gradient approximation (GGA) within the Perdew–Burke–Ernzerhof (PBE) parametrisation was used to describe the exchange and correlation functional. The projector–augmented wave (PAW) was used to model the interaction between valence electrons and ions plus core electrons, whereas we employed the ultrasoft method for the Fe atoms to lower the cut-off threshold for the wave function’ basis set. The energy (density) cutoff of the plane–wave expansion was set to 60 Ry (480 Ry), with a Gaussian smearing of 0.00735 Ry and a convergence threshold for the electronic self-consistent loop equal to 10$^{-6}$ Ry, while we sampled the Brillouin zone with a $\mathbf{k}$\nobreakdash-spacing of 0.1 \r{A}$^{-1}$. To take into account van der Waals dispersive forces, we used the DFT-D3 approach developed by Grimme et al.~\cite{ref:Grimme_2010}.

\subsection{Molecular systems and large-scale simulations set up}
\label{sec:mol_system}

\begin{figure}[htpb]
\centering
\includegraphics[width=0.7\columnwidth]{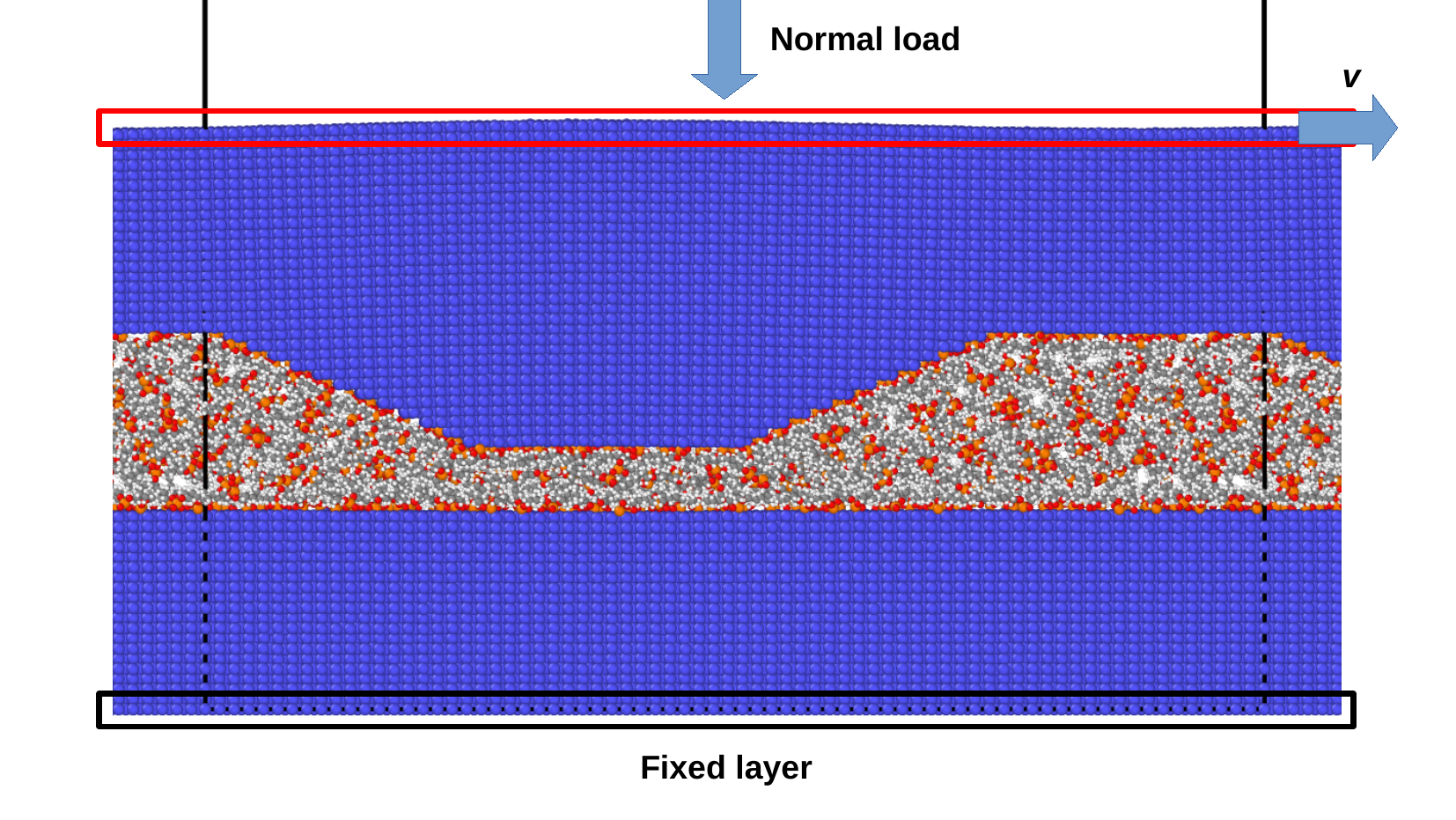}
\caption{\label{fig:Fe-pyr} Ball-and-stick representation of the computational setup for the system containing DBHP molecules confined within the iron interface (represented in blue). The black solid line represents the adopted simulation cell size. The red solid box represents the top-most iron layer where the normal load and the sliding velocity are applied, while the bottom-most black solid box identifies the reference iron layer, with fixed atomic positions.}
\end{figure}

We investigated three molecules for our molecular dynamics simulations, namely dibutyl hydrogen phosphite (DBHP), octyl acid phosphate (OAP) and methyl polyethylene glycol phosphate (mPEG-P) ester, as illustrated in Fig.~\ref{fig:mols}. As discussed in the Introduction, lubricant formulations typically contain a mixture of mono-, di-, and triester forms of mPEG-Ps, which were also included in our study (shown in Fig.~\ref{fig:esters}). The additives were confined between an asymmetric iron interface, composed of Fe(110) and Fe(211) surfaces, featuring a single asperity and an in-plane area of $20.93$ nm $\times$ $4.87$ nm, which is represented in Fig.~\ref{fig:Fe-pyr}. This setup, previously employed in tribological simulations~\cite{ref:Codrignani_2023}, assures the modelling of a more realistic, non-flat contact interface.

To simulate boundary lubrication conditions more realistically, we isolated the effect of lubricant additives by excluding the base oil from our simulations. This approach allows to focus on the behaviour of additives under sever boundary conditions, where the protective liquid lubricant mixture is assumed to be expelled from the contact and only the adsorbed molecules remain as friction-reducing agents. For each additive, we performed preliminary calculations to determine the amount of additives necessary to prevent the interfacial cold welding and eliminate voids within the interface. The optimal total molecular mass was approximately 132,800 atomic mass units, corresponding to 688 molecules of DBHP, 632 of OAP and 461 of the monoester mPEG-P. This setup resulted in simulation cells containing approximately 100,000 atoms across all systems.

All the large-scale molecular dynamics simulations were carried out using the LAMMPS package~\cite{ref:LAMMPS_2022} interfaced with DeePMD. We adopted a consistent simulation protocol across all system: the lubricant additives were initially placed randomly within the iron interface, followed by a system relaxation. The system was then linearly heated to the target temperature of 300 K using a timestep of 1 fs. After thermalisation, we configured the set up to model the non-equilibrium tribological interface under combined load and shear conditions. Specifically, we applied an NVT ensemble to the bulk iron layers and an NVE ensemble to the interfacial region. The bottommost layer of the lower iron surface was kept fixed, while a constant normal load of 1 GPa and a sliding velocity of 5 m/s were imposed on the topmost layer of the upper surface.

\subsection{Number of bonds per molecule (NBM) definition}

Due to the high number of dissociation events observed in our MD simulations, which is significantly larger than in our previous works~\cite{ref:Zilibotti_2013,ref:Loehlé_2018,ref:Restuccia_2020}, we introduced a descriptor called the number of bonds per molecule (NBM). This metric quantifies the average number of bonds for a specific type formed over time by a single molecule. For example, in the case of OAP, each molecule contains one phosphorus atom that forms four P--O bonds. The corresponding NBM for the P--O bond type is therefore 4, indicating that each molecule, on average, forms four of these bonds, as shown in Fig.~\ref{fig:mols}. In more complex cases, like the O--H bond in the OAP or mPEG-P where multiple oxygen and atoms are present, the NBM is equal to 2, reflecting the two O--H bonds typically found in each molecule.

The NBM is computed by multiplying the coordination number (CN) of a given bond type by the number of atoms for a specific atomic type. The CN is calculated by integrating the pair radial distribution function $g(r)$ up to the first minimum beyond the main peak, corresponding to the first coordination shell. It is defined mathematically as:

\begin{equation}
    CN = 4 \pi \int_{0}^{r_{min}} \rho \cdot g(r) \cdot r^{2} dr
\end{equation}

where $\rho$ is the atomic number density of the system, and $r_{min}$ is the position of the first minimum in the $g(r)$. This integration yields the average number of neighbouring atoms surrounding a reference atom within the first coordination shell. This approach allows to monitor the dissociation events of each bond type and to compare the distinct dissociative behaviours of the molecular additives under study.

\section{Results and discussion}

\subsection{Difference between phosphites and phosphates as friction modifiers}

\begin{figure}[htpb] 
    \centering 
    \includegraphics[width=0.7\columnwidth]{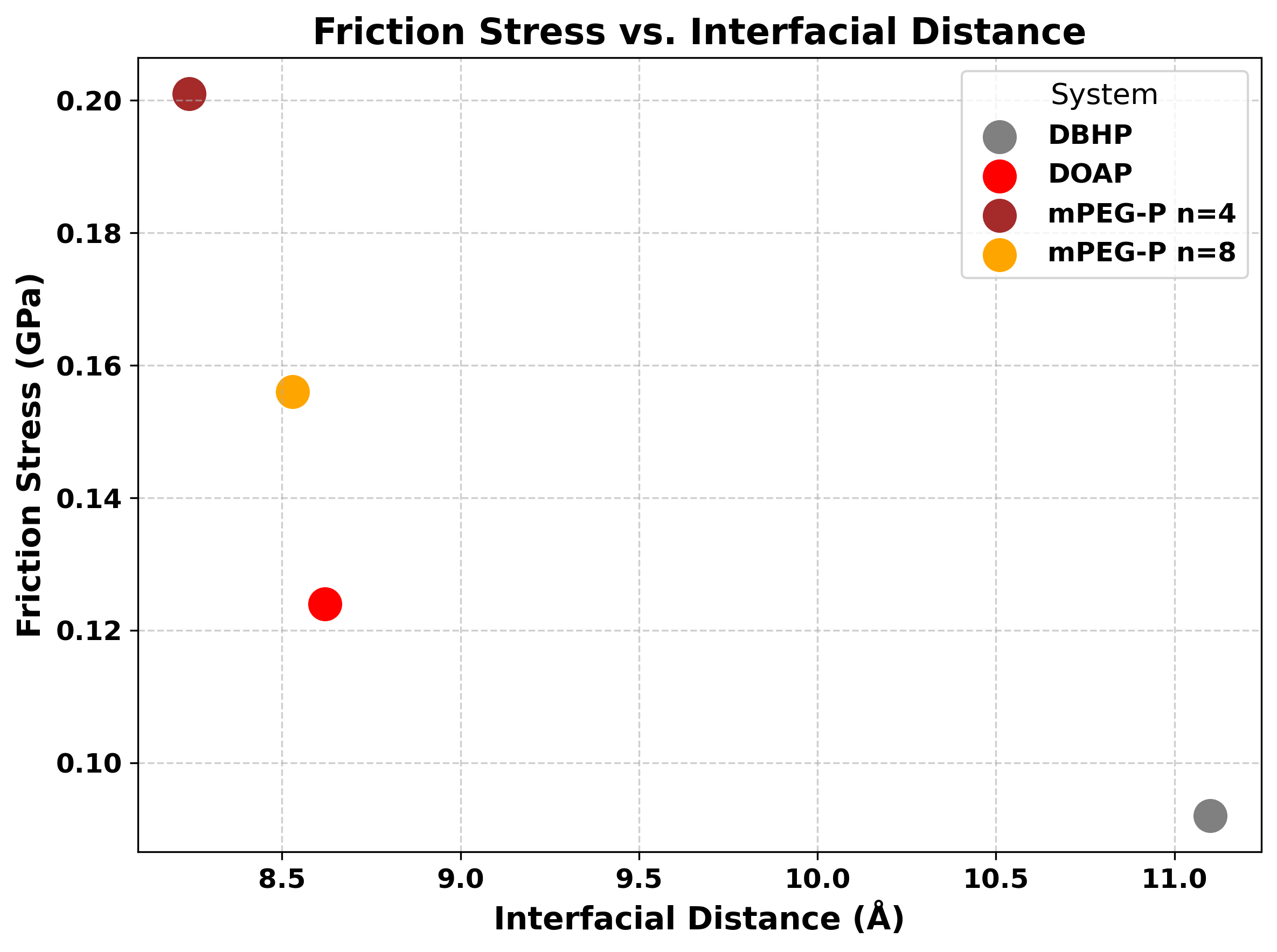}
    \caption{Friction vs distance for the systems containing mPEG-P monoester (n=4), mPEG-P monoester (n=8), OAP and DBHP.} 
    \label{fig:fri_v_dist} 
\end{figure}

After performing the sliding simulation of the Fe asperity, we evaluated the resistive force by computing the running average of the force acting on the sliding layer over the final 1 ns of the trajectory. This force was then compared to the interfacial separation between the asperity tip and the counter surface, a method previously employed to investigate the tribological behaviour of different additives~\cite{ref:Ta_2025}. As shown in Fig.~\ref{fig:fri_v_dist}, the interface lubricated by DBHP exhibits the largest interfacial separation and lowest friction among the tested additives. This inverse trend between boundary film thickness and friction is consistent with prior observations where tribofilm growth allows to lower friction~\cite{ref:Dawczyk_2019}. Notably, DBHP maintains a significantly larger separation (around 11 \r{A}) compared to the other P-based additives, which all cluster around 8.5 \r{A}). This result is a further confirmation of the lower friction for phosphites, which has been linked to the tribochemical formation of iron phosphide at ferrous interfaces, a phase that markedly reduces shear~\cite{DeBarros-Bouchet_2015}.

\begin{figure}[htpb] 
    \centering 
    \includegraphics[width=0.7\columnwidth]{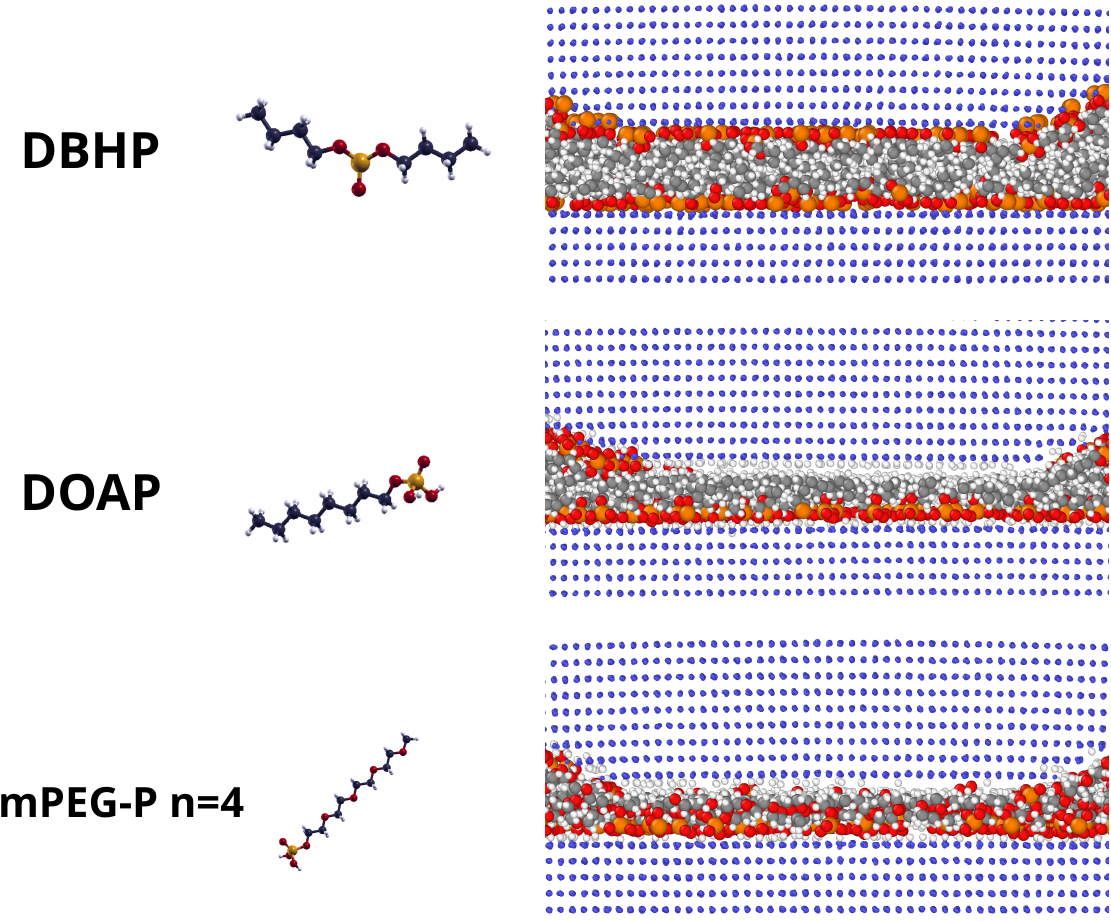}
    \caption{Ball-and-stick representation of the iron asperity containing DBHP (upper panel), OAP (central) and mPEG-P monoester with 4 glycol units (lower) additives at the end of the sliding simulations (19.5 ns). The dimension of Fe atoms has been reduced to make more visible the interaction of the P-based additives.} 
    \label{fig:MD_slide} 
\end{figure}

This significant difference in the tribological behaviour can be explained by the different tribochemical reactions occurring during the simulations. In particular, as shown in Fig.~\ref{fig:MD_slide}, the DBHP could adsorb uniformly over the asperity tip and the counter surface, whereas OAP and mPEG-P only covered the lower substrate of the simulation cell. Moreover, the DBHP alkyl chains seem to show a larger steric hindrance compared to OAP and mPEG-P additives, further increasing the interfacial distance with benefits in friction reduction. This qualitative behaviour is supported by quantitative data: DBHP exhibits higher molecular and dissociative adsorption energies, as reported in a recent study on phosphorus-based additives~\cite{ref:Benini_unpub}. More in general, strong chemisorption of organophosphorus additives on iron substrates is well documented in the literature, with surface reactions yielding polyphosphate-rich films (phosphates)~\cite{Latorre_2021} or P-rich species (phosphites)~\cite{ref:Gao_2004}. 

The analysis of the normalized bond metrics (NBMs) over time highlights key differences in chemical activity. Among the three additives, DBHP is the only one to exhibit significant P-–O bond dissociation (–16\% from the beginning to the end of the simulation). In contrast, both OAP and mPEG-P esters show high levels of O--H bond dissociation, associated with the loss of hydroxyl groups in the phosphate acid head group, approximately –50\% for OAP and –35\% for mPEG-P. Finally, all molecules have higher NBM for O--Fe bonds, but for different reasons: DBHP dissociates (increasing by 71\%), showing higher O--Fe bonds thanks to the freeing of central reactive units, whereas OAP (+78\%) and mPEG-P (+50\%) release H atoms and the O are free to bond to Fe. 

The marked differences in tribological behaviour among the additives can be attributed to their distinct tribochemical reactivities during sliding. As shown in Fig.~\ref{fig:MD_slide}, DBHP molecules adsorb uniformly across both the asperity tip and the counter surface, whereas OAP and mPEG-P primarily adhere to the lower substrate of the simulation cell. Additionally, the bulkier alkyl chains of DBHP appear to introduce greater steric hindrance compared to OAP and mPEG-P, contributing to an increased interfacial separation and, consequently, reduced friction.

All additives also display an increase in O--Fe bond formation, albeit via different mechanisms. For DBHP, this increase (+71\%) results from P--O dissociation, which exposes reactive oxygen atoms capable of bonding with Fe. In contrast, OAP (+78\%) and mPEG-P (+50\%) enhance O--Fe bonding primarily through hydrogen release, which leaves oxygen atoms free to coordinate with the iron surface.

\subsection{Different number of ester groups and chain length in mPEG-P}

\begin{figure}[htpb] 
    \centering 
    \includegraphics[width=0.7\columnwidth]{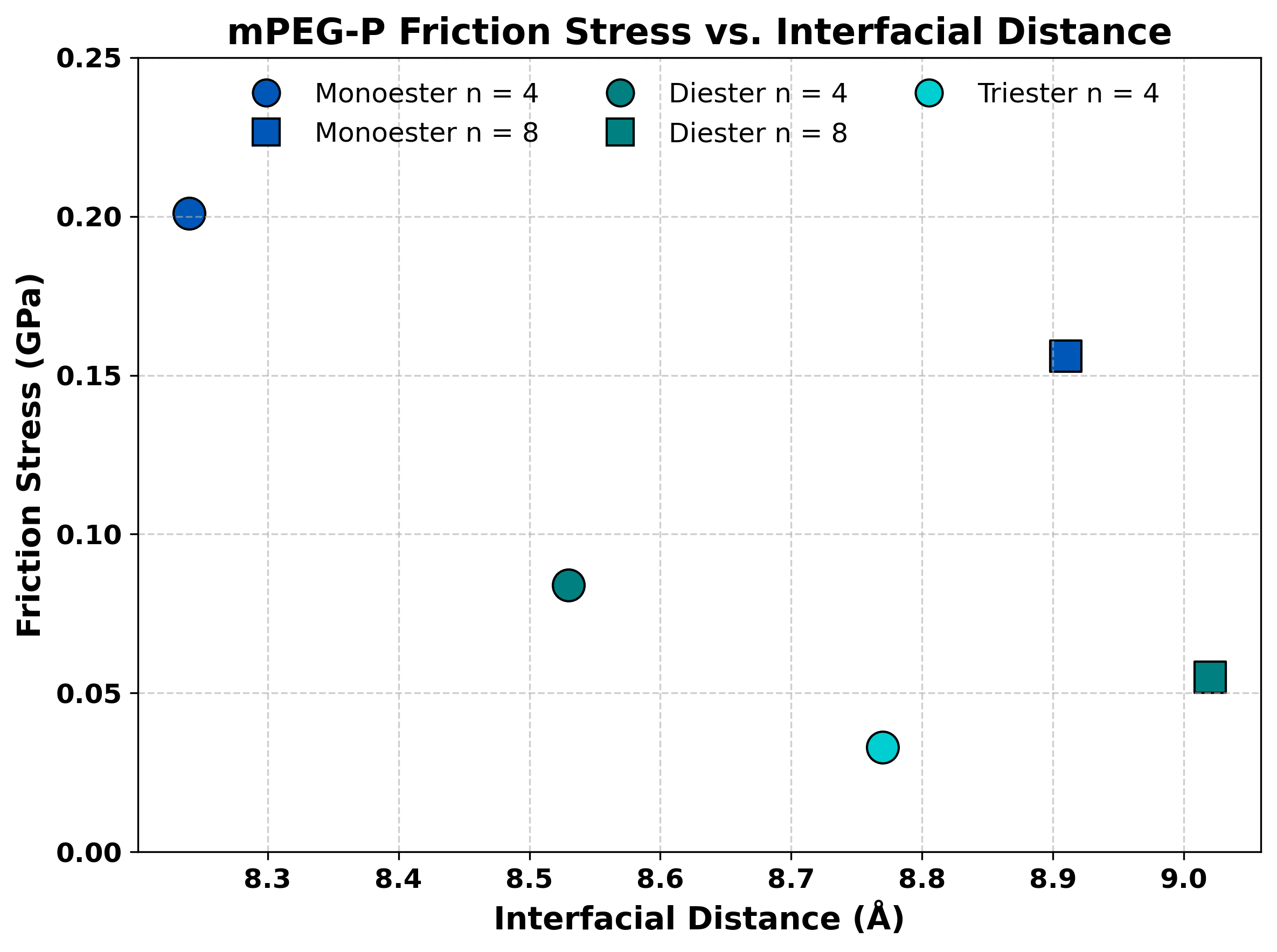}
    \caption{Friction vs distance for the systems containing mPEG-P monoester with 4 (8) glycol units (label A (D) in Fig.~\ref{fig:esters}), diester with 4 (8) glycol units (label B (E) in Fig.~\ref{fig:esters}) and triester with 4 glycol units (label C in Fig.~\ref{fig:esters}).} 
    \label{fig:fri_v_dist_Rhoda} 
\end{figure}

With the same simulation setup, we then focused on mPEG-P, studying the effect of the number of ester groups and chain length on their tribological performance. The resistive force (friction stress) and corresponding interfacial distance for each variant of mPEG-P in the last 1ns of sliding is reported in Figure~\ref{fig:fri_v_dist_Rhoda}. Two clear trends emerge: for fixed chain length, as the number of chains increases (going from monoester to triester), friction decreases and interfacial distance increases. The same trend is observed for fixed number of ester groups, as the chain length is increased. This suggests that the steric hindrance of the molecules plays a critical role in decreasing friction: larger amount of inert hydrocarbon chains provides improved slipperiness and load carrying capacity. Notably, increasing the number of ester groups for the same chain length provides a larger friction decrease with a smaller increase in interfacial distance, while increasing chain length for the same number of ester groups produces a larger interfacial separation but has a smaller impact on friction. This friction reduction with increasing chain number/length is consistent with polymer-brush steric lubrication: densely tethered chains resist compression, maintain separation and shear at low stress, thereby lowering the friction coefficient under boundary/mixed regimes~\cite{ref:Yang_2017,ref:Gmur_2021,ref:Abdelbar_2023}. Moreover, increasing ester functionality in base-oil esters thickens adsorbed films and reduces friction~\cite{ref:Hirata_2023}, confirming that multi-ester architectures promote separation and lower shear.

\begin{figure*}[htpb] 
    \centering 
    \includegraphics[width=\textwidth]{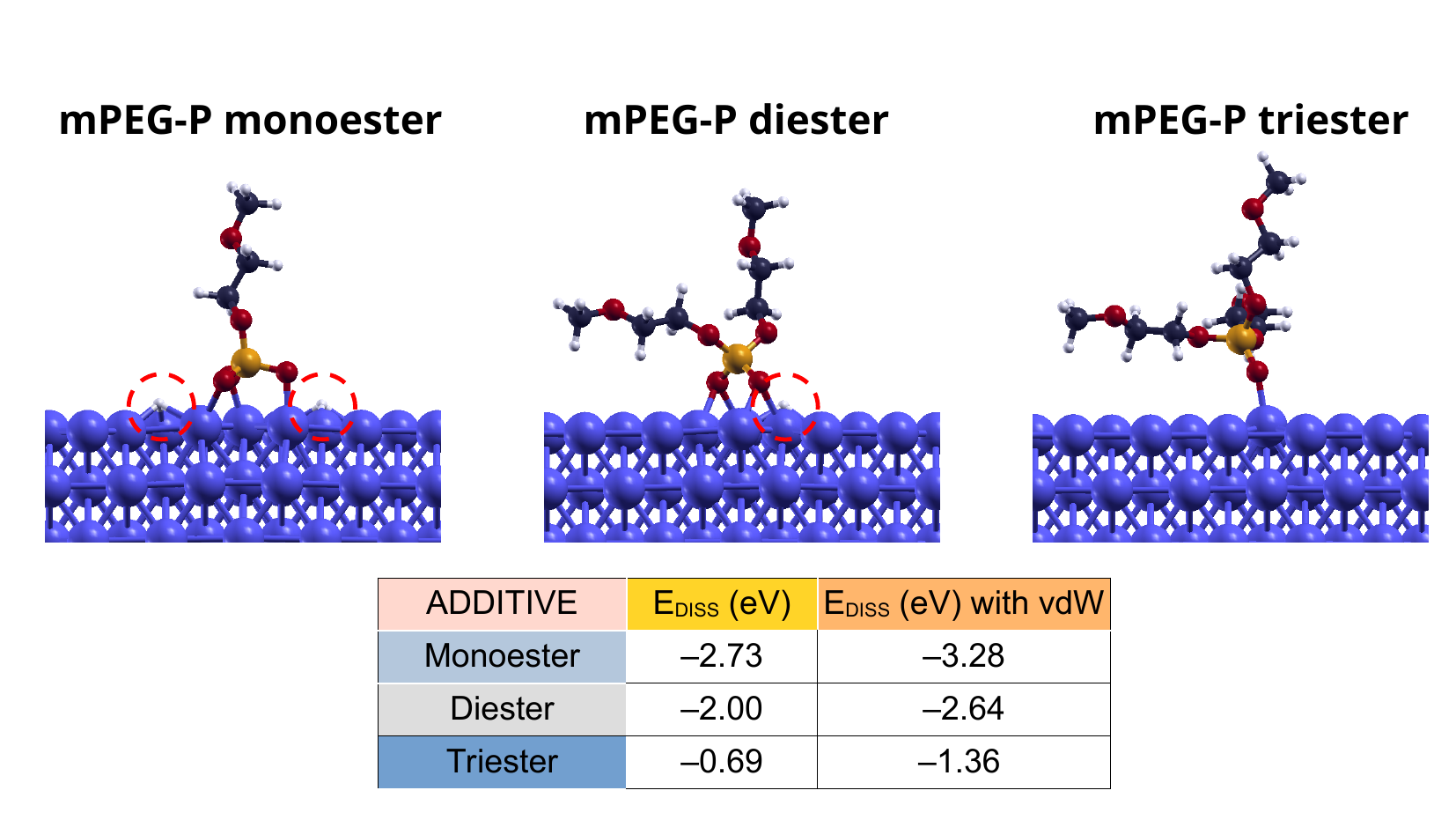}
    \caption{H- dissociative adsorption energy mPEG-P (n=1) mono-, di- and triester over a Fe(110) substrate.} 
    \label{fig:Rhoda_dissads} 
\end{figure*}

Interestingly, this trend in friction reduction contrasts with the adsorption behaviour of the mPEG-P molecules on the iron substrate. When computing the NBM variation for the O--Fe bonds, used as a proxy for the number of molecules attached to the surface, we found a sharper increase for mPEG-P monoesters (+50\% from beginning to end of the simulation) than for the diesters (+16\%) and triesters (+8\%). This result indicates that monoesters adhere more readily to the surface, despite exhibiting higher friction.

This trend is consistent with the O--H dissociative adsorption energies computed by DFT simulations, as shown in Fig.~\ref{fig:Rhoda_dissads}: the adsorption through the dissociation of the hydroxyl group in mPEG-P mono- and diesters has significantly higher energy gains, respectively by a factor 4 and 3, compared to the triester, which lacks O--H groups altogether. It is also well known that the head-group chemistry strongly modulates adsorption: monoesters with acidic O–H groups typically bind/react more readily on Fe/Fe-oxide than diesters/triesters~\cite{ref:Koshima_2024}. These insights further support the idea that steric hindrance in boundary lubrication can play a more dominant role than surface chemistry in friction reducing friction.

We did not consider the triester variant of mPEG-P with chain length n=8 in the comparison due to its large size, for two reasons. First, the molecule size is too large compared to the lateral dimensions of our simulation cell, which can produce artifacts due to periodic confinement. Second, the required amount of molecules to reach the same molecular weight as the other systems would be too small, below what can be considered a reasonable amount to be able to extract statistically significant results.

\section{Conclusions}

This work presents a comparative analysis of phosphorus-based lubricant additives, DBHP, OAP, and mPEG-P esters, under tribological conditions using machine learning-based molecular dynamics. The mPEG-P series was further investigated across different esterification degrees and chain lengths to disentangle the role of molecular architecture on tribological performance.

Our simulations reveal that DBHP exhibits the lowest friction and largest interfacial separation, a performance attributed to its enhanced steric hindrance and higher chemical reactivity. The bond analysis revealed substantial P--O bond dissociation and a marked increase in O--Fe bonding, indicating active tribochemical interactions that facilitate film formation and friction reduction. In contrast, OAP and mPEG-P monoesters produce thinner film and display higher friction due to reduced steric hindrance and lower resistance to shear displacements, which causes a decrease in surface coverage on contact asperities in extreme conditions.

Within the mPEG-P series, friction systematically decreases with both increasing esterification degree and chain length. In particular, the esterification degree exerts the strongest influence: di- and triesters species lower friction by providing a larger steric hindrance at the interface. Despite their weaker adsorption over metallic substrates, confirmed by the smaller increase in O--Fe bonds in MD simulations and lower O–H dissociation energies from static DFT calculations, these bulkier species maintain larger separations, consistent with a steric-dominant lubrication mechanism.

Overall, our findings highlight that under extreme confinement, steric hindrance and molecular architecture can outweigh surface reactivity in determining lubricant performance. The combination of machine learning-based MD and bond-metric analysis provides atomistic insight into the balance between adsorption, decomposition, and steric effects. These results outline rational design principles for next-generation phosphorus-based additives, suggesting that optimal formulations should balance reactive anchoring groups with sufficient molecular bulk to ensure low friction and durable protection under severe tribological conditions.

\section*{CRediT authorship contribution statement}
\textbf{Paolo Restuccia}: Formal analysis, Investigation, 
Writing – original draft. \textbf{Enrico Pedretti}: Formal analysis, Investigation, Writing – review \& editing. \textbf{Francesca Benini}: Investigation, Methodology. \textbf{Sophie Loehlé}: Conceptualization, Funding acquisition, Writing – review \& editing. \textbf{M. Clelia Righi}: Conceptualization, Funding acquisition, Supervision, Writing – review \& editing.

\section*{Acknowledgements}
MCR, PR, EP, and FB acknowledge the SLIDE project, which received funding from the European Research Council (ERC) under the European Union’s Horizon 2020 research and innovation program. (Grant Agreement No. 865633). We also acknowledge the CINECA award under the ISCRA initiative, for the availability of high-performance computing resources and support.

\section*{Data availability}
Data will be made available on request.

\bibliographystyle{elsarticle-num}
\bibliography{biblio}

\end{document}